\documentclass[10pt, conference, letterpaper]{IEEEtran}

\usepackage{amsmath,amssymb}
\usepackage{amsmath} 
\usepackage{graphicx}
\usepackage{subfigure}
\usepackage{url}
\usepackage{color}
\newcommand{\BfPara}[1]{{\noindent\bf#1.}\xspace}
\usepackage{multirow}

\usepackage{tikz}

\usepackage{epsfig}

\usepackage{xspace}

\definecolor{linkcolour}{rgb}{0,0.2,0.6}
\definecolor{xgreen}{rgb}{0.2,0.6,0.0}
\definecolor{xred}{rgb}{0.7,0.1,0.0}
\def\equationautorefname~#1\null{(#1)\null}

\colorlet{punct}{red!60!black}
\definecolor{background}{HTML}{ffffff }
\definecolor{delim}{RGB}{20,105,176}
\colorlet{numb}{magenta!60!black}

\definecolor{light-gray}{gray}{0.95}
\definecolor{darkgray}{rgb}{0.4, 0.4, 0.4}
\definecolor{editorGray}{rgb}{0.95, 0.95, 0.95}
\definecolor{editorOcher}{rgb}{1, 0.5, 0} 
\definecolor{editorGreen}{rgb}{0, 0.5, 0} 
\definecolor{orange}{rgb}{1,0.45,0.13}      
\definecolor{olive}{rgb}{0.17,0.59,0.20}
\definecolor{brown}{rgb}{0.69,0.31,0.31}
\definecolor{purple}{rgb}{0.38,0.18,0.81}
\definecolor{lightblue}{rgb}{0.1,0.57,0.7}
\definecolor{lightred}{rgb}{1,0.4,0.5}
\usepackage{upquote}
\usepackage{listings}

\definecolor{pblue}{rgb}{0.13,0.13,1}
\definecolor{pgreen}{rgb}{0,0.5,0}
\definecolor{pred}{rgb}{0.9,0,0}
\definecolor{pgrey}{rgb}{0.46,0.45,0.48}

\begin{document}

\title{Computer Systems Have 99 Problems, Let'€™s Not Make Machine Learning Another One}

\author{\IEEEauthorblockN{David Mohaisen$^\dagger$ and Songqing Chen$^\ddagger$}\\
$^\dagger$University of Central Florida \hspace{3mm} $^\ddagger$George Mason University}

\maketitle

\begin{abstract}
Machine learning techniques are finding many applications in computer systems, including many tasks that require decision making: network optimization, quality of service assurance, and security. We believe machine learning systems are here to stay, and to materialize on their potential we advocate a fresh look at various key issues that need further attention, including security as a requirement and system complexity, and how machine learning systems affect them. We also discuss reproducibility as a key requirement for sustainable machine learning systems, and leads to pursuing it.  
\end{abstract}

\IEEEpeerreviewmaketitle

\section{Introduction}\label{sec:introduction}
Over the past decade, machine learning has grown as scientific and perhaps independent discipline for the study of statistical models that can help in performing tasks, such as classification, without relying on explicitly defined rules, and rather on patterns and inferences made from data used for building those ``data models''.  For example, samples of data are typically used in machine learning algorithms to build a model in a process called ``model training'', which (the model) then is used for making decisions concerning samples of the same type as those used in building the model. 

The applications of machine learning in computer and networked systems are ubiquitous, with hundreds of applications and thousands of publications emerging every year. Those applications include areas of computer networks and simulation~\cite{ye2008large}, pattern recognition~\cite{bishop2006pattern}, general designs~\cite{ahn2018shmcaffe}, data quality assurance~\cite{lease2011quality}, attack detection and security enhancement~\cite{AlasmaryKAPCAAN19,SpauldingPKNM19,0002MC19,AbusnainaKA0AM19}, network and computer systems optimization through forecasting and classification~\cite{jordan2015machine}, user behavior analysis for prioritization and optimization~\cite{webb2001machine,kwon2016crime}, among many others. Machine learning techniques, including shallow and deep learning, have been widely utilized to problems in the computer and networked systems space, covering  fundamental problems such as congestion control~\cite{park2009intelligent}, traffic management~\cite{abuhmed2008survey}, resource management~\cite{martinez2009dynamic}, and quality of service (QoS) management and assurance~\cite{arslan2015training}. 

Some might argue that the use of machine learning in many of those applications is unjustifiable, as is becoming apparent in various of our community gatherings, arguing that the use of machine learning is driven by the general hype surrounding an emerging research area, and perhaps other alternatives and existing approaches should be utilized, including model-based (but not learning) algorithms that have been widely utilized and shown to provide good results in the literature. Some others have painted a different reality: machine learning is the only approach ahead for most if not all problems.

In this paper, we take a middle ground: machine learning is here to stay, and machine learning applications in many computer systems in general and networked systems in particular 
are perhaps justified, important, and necessary. However, in order for a meaningful use of machine learning in this area to materialize, various problems are certain, and addressing those problems as a first step requires a significant attention and effort from the community. More so, those problems require a fresh look, for that they are forever new with the particulars of machine learning systems. To instantiate our position, we consider three dimensions: security, complexity, and reproducibility. We highlight facets of those dimensions that require attention so that the introduction, integration, and utilization of machine learning algorithms into computer and networked systems are useful and usable.

\BfPara{Organization} 
The rest of this paper is organized as follows. In section~\ref{sec:security}, we address security of machine learning systems. In section~\ref{sec:complexity}, we discuss various issues with the inherent complexity of machine learning systems, and how they might affect the operation of computer systems. In section~\ref{sec:reprod} we draw attention to reproducibility, a central issue affecting the sustainability of this domain and our views on how to partly address it. Section~\ref{sec:conclusion} has concluding remarks.

\section{Security Is Not A Feature}\label{sec:security}
For long time, security and privacy have been considered as a feature, and not as a core computer system requirement, except when the computer system itself is built around security (e.g., authentication systems). For example, in networked systems, Distributed Denial of Service (DDoS) attacks are still prevalent today~\cite{WangMC17,WangMCC15} in great part because the Internet was first designed in principle as a medium of communication between endpoints, taking into account the reliability of the communication, and without any consideration for source authentication. It was envisioned back then that security would be addressed at the application layer, or incrementally by employing upgrades to existing protocols. 

\subsection{Time for Experiential Learning: A Prologue}
The past 30 years are a testimony of the failure of this argument. With a few exceptions, upgrades to insecure protocols, such as IP, DNS, and BGP, to add security features through alternatives, such as IPSec, DNSSEC, and BGPSec, have been fiercely resisted for many reasons, including overhead (e.g., computation overhead for signature verification) and trust (e.g., key certification). Where those new upgrades are not resisted, the shift to utilizing them has been very slow for the lack of incentives: utilizing those protocols in some cases means utilizing new stacks of hardware, by replacing existing infrastructure, which is quite expensive, with very little return if such upgrades are to be done eventually.  

Although not clearly visible yet, we expect the same scenario to repeat itself with machine learning algorithms being advocated today for every application. A natural question that might become apparent in this context is the following: why would we need to have security measures and an understanding of the security capabilities of a machine learning algorithm employed in a non-security application, e.g., resource management, queuing, quality of service, etc.? In other words, what are the faces of similarity, if any, between machine learning algorithms employed in computer systems and those of Internet protocols? The answer is multi-fold. 

First, while not apparent, security--broadly defined--is arguably part of those applications' basic guarantees. For example, one possible expected non-security guarantee of certain queuing algorithms is fairness. Allowing such an algorithm through machine learning to efficiently and reliably ensure fairness, while not explicitly stated as a security requirement, as a core security requirement. Having a secure machine learning algorithm, where the operating of the algorithm will help ensure the final outcome of the application in which it is employed, is essentially a requirement, not a feature. 

Second, we believe that machine learning needs to be viewed as part of the systems in which they are employed, and not as a stand-alone and separate component that does not influence the system. That is in fact the modus operandi in many systems and their design utilizing machine learning. To that end, a simple modification in the machine learning algorithm might affect the performance of the system, causing, among other outcomes, a violation of the very  basic guarantees of overall computer system. As demonstrated in the machine learning literature, this change can be in the algorithm itself or the input to the algorithm. Eventually, machine learning algorithms are implemented as a software; e.g., as in the widely used machine learning libraries, such as TensorFlow, Keras, Theano, PyTorch, and Pandas, among many others.  Modifying the algorithms could be as simple as exploiting a vulnerability (e.g., buffer overflow). Worse yet, machine learning algorithms are prone to adversarial examples: examples are are not necessarily out of the training dataset, but that once are slightly modified (or perturbed) they could be used to cause misclassification by the machine learning algorithm (e.g., labeling a malicious sample as benign). 

Third, the community has been advocating translating machine learning algorithms used in systems into hardware for performance reasons. Implementing those algorithms in hardware, without having a clear understanding of their security properties, nor of their resilience to manipulated input will eventually raise the cost of reversal: employing a more secure machine learning algorithm will require replacing the hardware implementation of insecure machine learning algorithms. 

Based on that, we envision various directions that are worthwhile, including the following.

\subsection{On Defining Realistic Threat Models}
Understanding the security of machine learning algorithms is only possible once we have a clear formulation of a threat model (or models) outlining the capabilities and adversarial goals. As a security community, we are very obsessed with threat modeling, and less so for the machine learning and systems community as the  community does not seem to agree on the terms.  Having those models is not an intellectual exercises, but is of paramount importance for analysis, evaluation, and comparison of systems. Moreover, standardized threat models in a way would help indirectly affirm capabilities of different systems based on the model they are analyzed within (as a framework).  A direct translation of threat models used for the abstract analysis of machine learning algorithms 
is plausible for that they are often theoretical in assuming unbounded capabilities. Realistically defining and standardizing threat models for machine learning systems would be a natural step drawing on their use model. 

A starting point towards this task would be by benefiting from the existing literature on machine learning systems and threat modeling in the adversarial machine learning community, and by outlining -- for a class of systems -- {\bf capabilities} of users and adversaries, and objectives of the adversaries. Given the decision making nature of most applications in which machine learning is utilized, those objectives can be then characterized, for example, as an intent to reduce confidence in this decision, to lead to a totally wrong but arbitrary decision (non-targeted misclassification), or to lead to a wrong decision of a particular choice (i.e., targeted misclassification). We note that the models, while might be general and encompass multiple applications, the ramifications of outcome might be application-specific. For example, with confidence reduction being the objective of an adversary, the machine learning model's output might not necessarily be a wrong outcome (e.g., imagine that the confidence of the model itself is being used as the model outcome for classification using thresholding). In such a case, understanding the operation of the system will entirely change how realistic, significant, or even meaningful is an attack of such nature. 

A second but equally important space for exploration would be capabilities associated with adversaries we will have to subject our machine learning systems to in order to analyze them. There has been a great deal of efforts in the learning (theory) community in general, and by the machine learning community in particular, in outlining capabilities into models, including white-box, gray-box, and black-box adversaries. Having a clear definition of those adversaries, especially in the context of machine learning systems as they are applied to various applications, would be essential. For example, while the white-box attack (whereby the adversary knows everything) and the black-box (whereby the adversary has an oracle-access to the model) are well understood, and are perhaps easy to translate to machine learning systems and applications, understanding which of those models is more relevant to a given system/application (or class of class of them) would be very important. With the potential difficultly of translating models across applications, it seems reasonable to annotate those broad classes of models into a hierarchy, and understand their relationship for different learning tasks, as those are going to eventually be of interest to analyze and contrast. Finally, while the gray model is perhaps figurative, it would be perhaps worthwhile to understand the degrees of gray associated with this model for different settings, applications, tasks, and contexts, and the relationships between them. Those definitions would perhaps benefit from outlining capabilities such as direct access to models and training data, oracle (or indirect) access to the model, and sample access (i.e., pairs of known sample input and the corresponding output). Using that as a broad model would a first step, but weighing in on how relevant each model is to a given application would certainly be essential, especially given the broad set of computer systems and applications in which they are employed.

\subsection{Understanding the Space of Robust Machine Learning}
Per Google scholar, there are more than 13,000 publications on adversarial machine learning in 2019 alone, not to mention that the year is not over yet (this text was written in November). In other words, it is not the lack of work on understanding the robustness of machine learning algorithms, but a) that few works exist on developing more robust machine learning algorithms that are easy to integrate in complex systems at reasonable cost, and b) few works that actually examine the robustness of machine learning algorithms in context. We highlight our concern with with both issues.

Examining machine learning algorithms in context (particularly in computer systems) is a key element that is lacking in the literature, and we (as a community) should pay more attention to that exploration. For example, there has been several studies in the literature on examining the robustness of (deep and shallow) machine learning algorithms against adversarial examples. In the context of computer vision, an ideal example to highlight the idea is the physical world attack on traffic signs: it was shown to be possible to launch a targeted attack making a detector (in a self-driving car, for example) detect the modified stop sign as a speed limit sign (for 45 mph)~\cite{eykholt2017robust}. While a clever attack, highlighting the shortcomings of deep learning networks and how easy it is to fool them by simple modifications, would only work in that context. In essence, the input sample from which the deep features are extract have no other purpose than the visual representation, recognized by a human, in a human-driven car, or by a sensor/classifier in self-driving cars. In most other computer systems, that is never the case: samples ingested by the machine learning algorithms embody an application-level and system operation logic that is often very complex, and inducing changes in the feature space would result in manipulating this logic. Such manipulation would essentially pronounce those adversarial examples useless. In the context of machine-learning based systems for anomaly detection, for example, modifying flows arbitrarily and independently may corrupt those flows, invaliding the key objective of the adversary: misclassification while keeping the sample intact~\cite{khormali2019examining}. 

Another example arise in the context of software classification, where the objective of the adversary is to force the machine learning algorithm to misclassify a malicious software (malware) as a benign one, for example~\cite{liu2019atmpa,GrossePMBM17,ShenVMKZ19}. In that context, there has been several works that provide involved approaches for such a task. Some of those approaches, for example, start with the feature space (a vector representation in the frequency domain; e.g., counts of bytes, header information, and flags), add some noise on the features, and force the machine learning system to misclassify the sample for which the features are extracted into malicious. This approach indeed works, except that it makes the strong assumption that the adversary has access to the pipeline of the operation of the machine learning algorithm/system~\cite{abs-1909-09735}. In other words, the approach makes the strong assumption that the adversary will be able to inject such noise in the pipeline at a stage between the actual input of a functional malware (or benign) sample and the execution of the machine learning algorithm on the abstract representation of this sample into a feature space. This is an unrealistic assumption. 

To relax this assumption, one can perhaps advocate one of two approaches: assuming a white-box model of an attack, one can perhaps pursue a backpropagation approach, whereby the noise itself is backpropagated from the feature space to the sample representation, or a forward-propagation whereby some modifications are introduced in the sample to generate a feature representation of interest (that is, a feature representation the results in the misclassification). While that would only relax the attack model, again, it would not necessarily guarantee a functioning sample. The backpropagation approach, although might be impractical and expensive, may as well corrupt the sample. On the other hand, the forward propagation approach, while could possibly result in a functioning sample, it may not result in a misclassification (given the constraints on the perturbation in the raw sample space). 

All in all, those issues call for further investigation into new metrics beyond misclassification, particularly metrics that are driven from the computer system application space to explain the results; e.g., executability, consistency, quality, etc.

\section{Watch for the Complexity}\label{sec:complexity}
As aforementioned, machine learning has been considered as an essential component in many systems and applications and is being considered for solving many problems. However, complexity is an intrinsic feature of many of those machine learning algorithms~\cite{chen2015mxnet}, making it difficult to imagine how such algorithms would fit in such systems without interfering with their basic properties if they could fit at all.

\BfPara{Model Complexity} Deep learning algorithms---such as artificial neural networks (ANN), recurrent neural networks (DNN), long short-term memory (LSTM), and convolutional neural networks (CNN)---are proposed to aid the security or operation (as an optimization subsystem) of various systems, including enterprise networks, Internet of Things (IoT) networks~\cite{berkayLeoIEEEspMagazine2019,SikderIEEETMC,BurakSOTA2018}, Mixed Reality (MR) systems~\cite{elbamby2018toward}, etc. Models of those machine learning systems are determined by the number of parameters, and are often in the order of thousands of parameters, amounting to 1-1000 Megabyte (e.g., the size allowed by the Amazon Machine Learning (AML) platform). 

While some of those computer systems might be appropriate for hosting those models (e.g., enterprise~\cite{berlin2015malicious}), some others might not. For example, such machine learning systems are unimaginable in the context of an on-device model in a home network system (in the IoT application). Such restriction calls for optimizations that should take into account reducing the complexity of models to a minimal size, and perhaps offloading the model on a dedicated hardware, thus indirectly increasing the complexity of the system. Introducing such components in the system, as a necessity due to introducing the machine learning component, would affect multiple aspects of the system, not the least of which the attack surface. 

\BfPara{Latency, Software, and Hardware Footprints} An additional concern to address is how to integrate such complex machine learning subsystems into a computer system pipeline. A majority of the literature on machine learning systems considers those systems in isolation, assuming a straightforward integration, which is rarely the case. We believe that particular considerations when designing machine learning techniques for computer systems should include, among other things, latency, software footprint, and hardware footprint. 
\begin{itemize}
    \item {\bf Latency} Networked systems designs nowadays aim to reduce latency to sub-10 millisecond margins on the Internet, often for application constraints. For example, in MR/VR applications, having an end-to-end latency of more than 15 milliseconds would pronounce such systems unusable (i.e., such a latency would result in poor experience, causing some users dizziness). The same phenomenon is present in multiple applications, and is not limited to MR/VR. To ensure such an end-to-end latency, the impact of the machine learning algorithms needs to be minimized. It is unclear, for example, with such large models and latency constraints, how to integrate machine learning algorithms in certain computer systems with such a low latency. It is our prediction that machine learning algorithms, eventually, would not be a viable solution for that  restriction alone. 
    
    \item {\bf Software and Hardware Footprint}  While storing the model would not require a lot of code representation (since using the model would be as simple as repeated multiplication operations, for example), it is sometimes beneficial to store the training code in the pipeline of the machine learning subsystem, e.g., for model retraining. Such a model, if made available in the system, needs to be integrated with system code (e.g., network buffers in the case of network optimization applications). Having a complex code would potentially expose the system in its entirety to vulnerabilities, and  call for reducing the code to the minimal functional components, while utilizing best practices and approaches for ensuring its security and reliability. Similarly, where hardware implementation is pursued for implementing machine learning subsystems, similar attention needs to be paid to how  the increase in complexity affects the operation of the overall system, including energy consumption (see section~\ref{sec:security} for more details on how security needs to be considered as a first citizen to reduce the cost of upgrades). 
    
\end{itemize}

\section{Addressing the Reproducibility Crisis}\label{sec:reprod}
In 2016, more than 1,500 scientists (including engineers;  {\url{https://tinyurl.com/y7bdswpj}} were surveyed on the reproducibility of scientific work and more than 70\% of them have pointed out that they failed to reproduce another scientist's experiments. What is more alarming is that 52\% of the surveyed scientists pointed out that they failed to reproduce their own work! Machine learning is no exception, either. A recent study surveying 30 research papers on machine learning approaches for texting mining has found that 27 (90\%) of them did not provide the raw data, 13 (43\%) did not provide sufficient details on data pre-processing, 9 (30\%) did not provide sufficient details on the feature representation and 19 (63\%) did not provide sufficient details on feature selection. Almost all of them did not provide code, 12 (40\%) did not provide algorithm details, and 15 (50\%) did not provide an executable file. All of them did not provide link to data partitions, data indices, or seed values, while 29 (97\%) of them did not provide appropriate version of the software used. The list goes on and on, and the state of affair in the world of the reproducibility is quite ugly. 

In major machine learning and system conferences, various initiatives have emerged around artifacts collection and evaluation, whereby a committee of selected researchers is tasked with checking the availability of artifacts, and evaluating them as functional or reusable. Upon that, a badge of available artifacts, function, or reusable is assigned to the work. However, to the best of our knowledge, those efforts are still optional, and a small minority of authors opt-out to submitting those artifacts. While a step in the right direction, sharing those artifacts with the evaluating committee a) does not necessarily mean that the results are reproducible, and b) does not mean that the artifacts would be accessible to the community beyond the evaluation committee or the evaluation time. 

In computer systems, as highlighted earlier, reproducibility depends on a lot of interdependent pieces, and not only the machine learning algorithms developed for a certain task. For example, performance evaluation and validation taking into account model complexity, measured by the time it takes to train or test the model, would depend on the used hardware, optimization parameters, settings, etc., and having those ``parameters'' standardized would be essential. Moreover, for a class of applications (e.g., QoS, QoE, energy optimization, anomaly detection, etc.), having standard traces addressing various aspects of the computer system and representative of the various use scenarios would be essential.  Data alone is not sufficient to addressing reproducibility, but essential. Other components include software, configurations, and libraries. One approach to facilitating reproducibility would to share those settings as lightweight containers or fully-fledged virtual machines with those installed in them. It is ideal to share the machine learning system in a form or another. For example, when developed as a software, the code should be made accessible, and if not accessible, perhaps a detailed description of the main building block should be made available for reproducibility. Even when not being able to share their code with others for reproducing results (e.g., for intellectual property or privacy constraints), they can still aid reproducibility by running their code/system on standardized settings and benchmark datasets. 

We believe addressing reproducibility is in an essence a core requirement for the sustainability of scientific endeavor, and a close attention must be paid to facilitating this requirement by ensuring that machine learning systems are shared in the public space in standard format that improve reproducibility.

\section{Concluding Remarks}\label{sec:conclusion}
Machine learning systems are here to stay. They might not be the solution to every problem in computer systems, given the heterogeneity of those systems and their requirements, but they promise various capabilities and optimizations in many others, thus are beneficial. To materialize the potential of machine learning systems, and not add to an already complex systems space, we advocate that more effort needs to be paid to at least three directions in the machine learning systems community: security, complexity, and reproducibility. 

\BfPara{Acknowledgement} This work was supported in part by a NRF grant 2016K1A1A2912757, a NSF grant CNS-1524462, and a NIST grant 70NANB18H272.



\begin{thebibliography}{10}
\providecommand{\url}[1]{#1}
\csname url@samestyle\endcsname
\providecommand{\newblock}{\relax}
\providecommand{\bibinfo}[2]{#2}
\providecommand{\BIBentrySTDinterwordspacing}{\spaceskip=0pt\relax}
\providecommand{\BIBentryALTinterwordstretchfactor}{4}
\providecommand{\BIBentryALTinterwordspacing}{\spaceskip=\fontdimen2\font plus
\BIBentryALTinterwordstretchfactor\fontdimen3\font minus
  \fontdimen4\font\relax}
\providecommand{\BIBforeignlanguage}[2]{{%
\expandafter\ifx\csname l@#1\endcsname\relax
\typeout{** WARNING: IEEEtran.bst: No hyphenation pattern has been}%
\typeout{** loaded for the language `#1'. Using the pattern for}%
\typeout{** the default language instead.}%
\else
\language=\csname l@#1\endcsname
\fi
#2}}
\providecommand{\BIBdecl}{\relax}
\BIBdecl

\bibitem{ye2008large}
T.~Ye, H.~T. Kaur, S.~Kalyanaraman, and M.~Yuksel, ``Large-scale network
  parameter configuration using an on-line simulation framework,''
  \emph{IEEE/ACM Transactions on Networking (TON)}, vol.~16, no.~4, pp.
  777--790, 2008.

\bibitem{bishop2006pattern}
C.~M. Bishop, \emph{Pattern recognition and machine learning}.\hskip 1em plus
  0.5em minus 0.4em\relax Springer Science+ Business Media, 2006.

\bibitem{ahn2018shmcaffe}
S.~Ahn, J.~Kim, E.~Lim, W.~Choi, A.~Mohaisen, and S.~Kang, ``Shmcaffe: A
  distributed deep learning platform with shared memory buffer for hpc
  architecture,'' in \emph{2018 IEEE 38th International Conference on
  Distributed Computing Systems (ICDCS)}.\hskip 1em plus 0.5em minus
  0.4em\relax IEEE, 2018, pp. 1118--1128.

\bibitem{lease2011quality}
M.~Lease, ``On quality control and machine learning in crowdsourcing,'' in
  \emph{Workshops at the Twenty-Fifth AAAI Conference on Artificial
  Intelligence}, 2011.

\bibitem{AlasmaryKAPCAAN19}
\BIBentryALTinterwordspacing
H.~Alasmary, A.~Khormali, A.~Anwar, J.~Park, J.~Choi, A.~Abusnaina, A.~Awad,
  D.~Nyang, and A.~Mohaisen, ``Analyzing and detecting emerging internet of
  things malware: {A} graph-based approach,'' \emph{{IEEE} Internet of Things
  Journal}, vol.~6, no.~5, pp. 8977--8988, 2019. [Online]. Available:
  \url{https://doi.org/10.1109/JIOT.2019.2925929}
\BIBentrySTDinterwordspacing

\bibitem{SpauldingPKNM19}
\BIBentryALTinterwordspacing
J.~Spaulding, J.~Park, J.~Kim, D.~Nyang, and A.~Mohaisen, ``Thriving on chaos:
  Proactive detection of command and control domains in internet of
  things-scale botnets using {DRIFT},'' \emph{Trans. Emerging
  Telecommunications Technologies}, vol.~30, no.~4, 2019. [Online]. Available:
  \url{https://doi.org/10.1002/ett.3505}
\BIBentrySTDinterwordspacing

\bibitem{0002MC19}
A.~Wang, A.~Mohaisen, and S.~Chen, ``{XLF:} {A} cross-layer framework to secure
  the internet of things (iot),'' in \emph{39th {IEEE} International Conference
  on Distributed Computing Systems, {ICDCS} 2019, Dallas, TX, USA, July 7-10,
  2019}, 2019, pp. 1830--1839.

\bibitem{AbusnainaKA0AM19}
A.~Abusnaina, A.~Khormali, H.~Alasmary, J.~Park, A.~Anwar, and A.~Mohaisen,
  ``Adversarial learning attacks on graph-based iot malware detection
  systems,'' in \emph{39th {IEEE} International Conference on Distributed
  Computing Systems, {ICDCS} 2019}, 2019, pp. 1296--1305.

\bibitem{jordan2015machine}
M.~I. Jordan and T.~M. Mitchell, ``Machine learning: Trends, perspectives, and
  prospects,'' \emph{Science}, vol. 349, no. 6245, pp. 255--260, 2015.

\bibitem{webb2001machine}
G.~I. Webb, M.~J. Pazzani, and D.~Billsus, ``Machine learning for user
  modeling,'' \emph{User modeling and user-adapted interaction}, vol.~11, no.
  1-2, pp. 19--29, 2001.

\bibitem{kwon2016crime}
H.~Kwon, A.~Mohaisen, J.~Woo, Y.~Kim, E.~Lee, and H.~K. Kim, ``Crime scene
  reconstruction: Online gold farming network analysis,'' \emph{IEEE
  Transactions on Information Forensics and Security}, vol.~12, no.~3, pp.
  544--556, 2016.

\bibitem{park2009intelligent}
J.~Park, Z.~Chen, L.~Kiliaris, M.~L. Kuang, M.~A. Masrur, A.~M. Phillips, and
  Y.~L. Murphey, ``Intelligent vehicle power control based on machine learning
  of optimal control parameters and prediction of road type and traffic
  congestion,'' \emph{IEEE Transactions on Vehicular Technology}, vol.~58,
  no.~9, pp. 4741--4756, 2009.

\bibitem{abuhmed2008survey}
T.~AbuHmed, A.~Mohaisen, and D.~Nyang, ``A survey on deep packet inspection for
  intrusion detection systems,'' \emph{arXiv preprint arXiv:0803.0037}, 2008.

\bibitem{martinez2009dynamic}
J.~F. Martinez and E.~Ipek, ``Dynamic multicore resource management: A machine
  learning approach,'' \emph{IEEE MICRO}, pp. 8--17, 2009.

\bibitem{arslan2015training}
E.~Arslan, M.~Yuksel, and M.~H. Gunes, ``Training network administrators in a
  game-like environment,'' \emph{Journal of Network and Computer Applications},
  vol.~53, pp. 14--23, 2015.

\bibitem{WangMC17}
A.~Wang, A.~Mohaisen, and S.~Chen, ``An adversary-centric behavior modeling of
  {DDoS} attacks,'' in \emph{37th {IEEE} International Conference on
  Distributed Computing Systems, {ICDCS} 2017}, 2017, pp. 1126--1136.

\bibitem{WangMCC15}
A.~Wang, A.~Mohaisen, W.~Chang, and S.~Chen, ``Capturing ddos attack dynamics
  behind the scenes,'' in \emph{Detection of Intrusions and Malware, and
  Vulnerability Assessment - 12th International Conference, {DIMVA} 2015,
  Milan, Italy, July 9-10, 2015, Proceedings}, 2015, pp. 205--215.

\bibitem{eykholt2017robust}
K.~Eykholt, I.~Evtimov, E.~Fernandes, B.~Li, A.~Rahmati, C.~Xiao, A.~Prakash,
  T.~Kohno, and D.~Song, ``Robust physical-world attacks on deep learning
  models,'' \emph{arXiv preprint arXiv:1707.08945}, 2017.

\bibitem{khormali2019examining}
A.~Abusnaina, A.~Khormali, D.~Nyang, M.~Yuksel, and A.~Mohaisen, ``Examining
  the robustness of learning-based ddos detection in software defined
  networks,'' in \emph{Proceedings of the IEEE conference on dependable and
  secure computing, IDSC}, 2019.

\bibitem{liu2019atmpa}
X.~Liu, J.~Zhang, Y.~Lin, and H.~Li, ``Atmpa: Attacking machine learning-based
  malware visualization detection methods via adversarial examples,'' 2019.

\bibitem{GrossePMBM17}
K.~Grosse, N.~Papernot, P.~Manoharan, M.~Backes, and P.~D. McDaniel,
  ``Adversarial examples for malware detection,'' in \emph{Proceedings of the
  22nd European Symposium on Research Computer Security - {ESORICS}, Part
  {II}}, 2017, pp. 62--79.

\bibitem{ShenVMKZ19}
\BIBentryALTinterwordspacing
F.~Shen, J.~D. Vecchio, A.~Mohaisen, S.~Y. Ko, and L.~Ziarek, ``Android malware
  detection using complex-flows,'' \emph{{IEEE} Trans. Mob. Comput.}, vol.~18,
  no.~6, pp. 1231--1245, 2019. [Online]. Available:
  \url{https://doi.org/10.1109/TMC.2018.2861405}
\BIBentrySTDinterwordspacing

\bibitem{abs-1909-09735}
\BIBentryALTinterwordspacing
A.~Khormali, A.~Abusnaina, S.~Chen, D.~Nyang, and A.~Mohaisen, ``{COPYCAT:}
  practical adversarial attacks on visualization-based malware detection,''
  \emph{CoRR}, vol. abs/1909.09735, 2019. [Online]. Available:
  \url{http://arxiv.org/abs/1909.09735}
\BIBentrySTDinterwordspacing

\bibitem{chen2015mxnet}
T.~Chen, M.~Li, Y.~Li, M.~Lin, N.~Wang, M.~Wang, T.~Xiao, B.~Xu, C.~Zhang, and
  Z.~Zhang, ``Mxnet: A flexible and efficient machine learning library for
  heterogeneous distributed systems,'' \emph{arXiv preprint arXiv:1512.01274},
  2015.

\bibitem{berkayLeoIEEEspMagazine2019}
Z.~B. {Celik}, P.~{McDaniel}, G.~{Tan}, L.~{Babun}, and A.~S. {Uluagac},
  ``Verifying internet of things safety and security in physical spaces,''
  \emph{IEEE Security Privacy}, vol.~17, no.~5, pp. 30--37, Sep. 2019.

\bibitem{SikderIEEETMC}
A.~K. {Sikder}, H.~{Aksu}, and A.~S. {Uluagac}, ``A context-aware framework for
  detecting sensor-based threats on smart devices,'' \emph{IEEE Transactions on
  Mobile Computing}, pp. 1--1, 2019.

\bibitem{BurakSOTA2018}
H.~B. {Yesilyurt}, H.~{Aksu}, S.~{Uluagac}, and R.~{Beyah}, ``Sota: Secure
  over-the-air programming of iot devices,'' in \emph{MILCOM 2018 - 2018 IEEE
  Military Communications Conference (MILCOM)}, Oct 2018, pp. 1--8.

\bibitem{elbamby2018toward}
M.~S. Elbamby, C.~Perfecto, M.~Bennis, and K.~Doppler, ``Toward low-latency and
  ultra-reliable virtual reality,'' \emph{IEEE Network}, vol.~32, no.~2, pp.
  78--84, 2018.

\bibitem{berlin2015malicious}
K.~Berlin, D.~Slater, and J.~Saxe, ``Malicious behavior detection using windows
  audit logs,'' in \emph{Proceedings of the 8th ACM Workshop on Artificial
  Intelligence and Security}.\hskip 1em plus 0.5em minus 0.4em\relax ACM, 2015,
  pp. 35--44.

\end{thebibliography}
\end{document}